\documentclass[%
onecolumn,%
oneside,%
floats,%
aps,%
 prd,%
nobibnotes,%
nofootinbib,%
amsmath,%
amssymb,%
amsfonts,%
amscd,%
  superscriptaddress,%
eqsecnum%
]{revtex4}

\usepackage[utf8]{inputenc}
\usepackage{graphicx,array,dcolumn}
\usepackage{cases}
\usepackage[paperwidth=210mm,paperheight=297mm,centering,hmargin=1.8cm,vmargin=2.5cm]{geometry}
\usepackage{enumerate}

\usepackage{hyperref}
\hypersetup{hidelinks}

\newcommand{\rar}{\rightarrow}

\def\be{\begin{eqnarray}}
\def\ee{\end{eqnarray}}

\def\-g{\sqrt{-g}}

\renewcommand\rho{\varrho}

\begin{document}


\title{Intrinsic problems of the gravitational baryogenesis }

\author{E.V. Arbuzova}
\email{arbuzova@uni-dubna.ru}
\affiliation{Novosibirsk State University, Novosibirsk, 630090, Russia}
\affiliation{Department of Higher Mathematics, Dubna State University, 141980 Dubna, Russia}

\author{A.D. Dolgov}
\email{dolgov@fe.infn.it}
\affiliation{Novosibirsk State University, Novosibirsk, 630090, Russia}
\affiliation{ITEP, Bol. Cheremushkinsaya ul., 25, 117259 Moscow, Russia}

\begin{abstract}

Modification of gravity due to the curvature dependent term in the gravitational baryogenesis scenario 
is considered. It is shown that this term leads to the fourth order  differential equation of motion for the 
curvature scalar instead of the algebraic one of General Relativity (GR). The fourth order gravitational 
equations are generically unstable with respect to small perturbations. Non-linear in curvature terms 
may stabilize the solution but the magnitude of the stabilized curvature scalar would be much larger than 
that dictated by GR, so  the standard cosmology would be strongly distorted.

\end{abstract}

\maketitle

\section{Introduction}

Different scenarios of baryogenesis are based, as a rule, on three well known Sakharov 
principles~\cite{ads}: \\
1. Non-conservation of baryonic number.\\
2. Breaking of symmetry between particles and antiparticles.\\
3. Deviation from thermal equilibrium.\\
For details see e.g. reviews~\cite{AD-BG,BG-rev}.

However, as is mentioned in review~\cite{AD-BG}, none of these conditions is obligatory.
{There  are versions of baryogenesis scenarios which can be successfully realized even if baryonic number, $B$, is strictly
conserved. For example, if there exists a hidden or sterile baryon, the baryon asymmetry in the sector of our
visible baryons would be exactly equal to the asymmetry in the invisible (or slightly visible) sector
consisting of hidden baryons. In particular,
such new baryons could form cosmological dark matter. Another example of baryogenesis with conserved $B$
is the model of generation of baryon asymmetry in the process of black hole evaporation. In this case the compensating
antibaryon asymmetry would be hidden inside the evaporating black holes. 

Out of the three Sakharov principles only non-conservation of baryons is necessary for 
spontaneous baryogenesis (SBG) in its classical version~\cite{spont-BG},
but this mechanism 
does not demand an explicit C and CP violation and  can proceed in thermal equilibrium. Moreover, it is usually most 
efficient in thermal equilibrium. 

The statement that the cosmological baryon asymmetry can be created by SBG in thermal equilibrium was 
mentioned in the original
papers~\cite{spont-BG}. The reaction rate may be very high, such that all essential processes are forced to equilibrium,
so the particle distributions do not deviate from  the canonical equilibrium form. Nevertheless in this thermal equilibrium
state an asymmetry between particles and antiparticles happened to be non-vanishing. Temporal evolution of the 
(pseudo)goldstone field $\theta (t)$, see below, in a sense mimics the non-equilibrium situation but this has nothing to do
with thermal equilibrium, it is simply induced by a deviation from the mechanical equilibrium point of $\theta (t)$.

As for C or CP violation, they are effectively present in SBG due to the external field $\theta (t)$, which acts differently on particles
and antiparticles but this field ultimately disappears and thus does not lead  to C and CP violation in particles physics
in our laboratories. It may be a matter of terminology, but we do not say e.g. that the different scattering of particles and 
antiparticles on a positively charged nucleus means C or CP violation.
}

The term "spontaneous" is related to spontaneous breaking of  a global $U(1)$-symmetry, which ensures the 
conservation of the total baryonic number in the unbroken phase. 
When the symmetry is broken, the baryonic
current becomes non-conserved and the Lagrangian density  acquires the term
\be
{\cal L}_{SB} =  (\partial_{\mu} \theta) J^{\mu}_B\, ,
 \label{L-SB}
 \ee
 where $\theta$ is the Goldstone (or pseudo Goldstone) field and $J^{\mu}_B$ is the baryonic current of matter fields. 

For a spatially homogeneous field $\theta = \theta (t)$  the Lagrangian (\ref{L-SB})
is reduced to ${\cal L}_{SB} =   \dot \theta\, n_B$, where $n_B\equiv J^0_B$ 
is the baryonic number density of matter, so it is tempting to identify $(-\dot \theta)$ with the 
baryonic chemical potential, $\mu_B$, of the corresponding 
system. The identification of $\dot\theta$ with chemical potential is questionable
and depends upon the representation chosen for the fermionic fields, as is argued in 
refs.~\cite{ad-kf,ADN}, but still the scenario is operative and presents a beautiful possibility
to create an excess of particles over antiparticles in the universe.

Subsequently the idea of gravitational baryogenesis (GBG) was put forward~\cite{GBG-1}, where the 
scenario of SBG was modified by the introduction of the coupling of the baryonic current to the derivative 
of the curvature scalar $R$:
\be
{\cal L}_{GBG} = \frac{1}{M^2} (\partial_\mu  R ) J^\mu_B\, ,
\label{L-GBG}
\ee 
where $M$ is a constant parameter with the dimension of mass. More references on GBG can be 
found in~\cite{GBG-more}.

In this work we demonstrate that the addition of the curvature dependent term (\ref{L-GBG}) to the 
Hilbert-Einstein Lagrangian of GR leads to higher order gravitational equations of motion which are strongly unstable with
respect to small perturbations. A similar gravitational instability in $F(R)$ gravity, which also leads to higher order 
equations of motion, has been found in refs.~\cite{grav-instab}.

\section{Equations of motion \label{sec-EoM}}

Let us start from the model where baryonic number is carried by scalar field $\phi$ with 
potential $U(\phi, \phi^* )$.
An example with baryonic current of fermions will be considered elsewhere.

The action of the scalar model has the form:
\be 
A = \int d^4 x\, \sqrt{-g} \left[ \frac{m_{Pl}^2}{16\pi } R + \frac{1}{M^2} (\partial_{\mu} R) J^{\mu}  - 
g^{\mu \nu} \partial_{\mu}\phi\, \partial_{\nu}\phi^* + U(\phi, \phi^*)\right] - A_m\, ,
\label{act-tot}
\ee
where $m_{Pl}=1.22\cdot 10^{19}$ GeV is the Planck mass, 
$A_m$ is the matter action,  $J^\mu = g^{\mu\nu}J_\nu$, and $g^{\mu\nu}$ is the metric tensor of the 
background space-time. We assume that initially the metric has the usual GR form and study the emergence 
of the corrections due to the  instability described below.

In contrast to scalar electrodynamics, the baryonic current of scalars is not uniquely defined. In electrodynamics 
the form of the electric current  is dictated by the conditions of gauge invariance and current conservation, which
demand the addition to the current 
of the so called sea-gull term proportional to $e^2 A_\mu |\phi|^2$, where $A_\mu$ is the electromagnetic potential.

On the other hand, a local $U(1)$-symmetry is not imposed on the theory determined by action (\ref{act-tot}).
It is invariant only with respect to  a $U(1)$ transformations with constant phase.
As a result,  the baryonic current of scalars
is considerably less restricted. In particular, we can add to the current an analogue of the sea-gull term, 
$ \sim (\partial_\mu R)\,|\phi |^2 $, with an arbitrary coefficient.
So we study the following two extreme possibilities, when the sea-gull term is absent 
{ and the current is not conserved
(the case A below), or the sea-gull term is included with the coefficient ensuring current conservation
(the case B). In both cases no baryon asymmetry can be generated without additional interactions. 
It is trivially true in the case B, when the current is conserved,
but it is also true in the case A despite 
the current non-conservation, simply because the non-zero divergence $D_\mu J^\mu$
does not change the baryonic number of $\phi$ but only leads to redistribution of particles $\phi$ in the phase space.
So to create any non-zero baryon asymmetry we have to introduce an interaction of $\phi$ with other particles 
which breaks  conservation of $B$ by making the potential $U$ non-invariant with respect to 
the phase rotations of $\phi$, as it is described below. }

\subsection{Current: version 1}
If the potential $U(\phi)$
is not invariant with respect to the $U(1)$-rotation, $\phi \rightarrow \exp{(i \beta)} \phi $, 
the baryonic current  defined in the usual way    
\be 
J_{1 \mu} = i q (\phi^* \partial_{\mu}\phi - \phi \partial_{\mu}\phi^*)
\label{current}
\ee
is not conserved. Here $q$ is the baryonic number of $\phi $ and
for brevity we omitted index $B$ in current $J_{1\mu}$.  

With this current and Lagrangian (\ref{L-GBG}) the equations 
for the gravitational field take the form:
\be \nonumber
&& \frac{m_{Pl}^2}{16\pi } \left( R_{\mu\nu} - \frac{1}{2} g_{\mu\nu} R \right) -
\frac{1}{M^2}\left( \left[R_{\mu \nu} - (D_{\mu}D_{\nu} - g_{\mu\nu}D^2)\right] D_{ \alpha} J^{\alpha}_{1 } +
\frac{1}{2} g_{\mu\nu} J^{\alpha}_1\,D_{\alpha} R  - \frac{1}{2} (J_{1 \nu} D_{\mu} R + J_{1 \mu} D_{\nu} R)\right)  \\ 
&&- \frac{1}{2} (D_{\mu} \phi \, D_{\nu} \phi^* + D_{\nu} \phi \, D_{\mu} \phi^*) + 
\frac{1}{2} g_{\mu\nu} ( D_{\alpha} \phi \, D^{\alpha} \phi^* -  U(\phi )) =  \frac{1}{2}\, T_{\mu\nu}\, ,
\label{EoM}
\ee
where $D_\mu$ is the covariant derivative in metric $g_{\mu\nu}$ (of course, for scalars $D_\mu = \partial_\mu$) and
$T_{\mu\nu}$ is the energy-momentum tensor of matter obtained from action $A_m$.  

Taking the trace of equation (\ref{EoM}) with respect to $\mu$ and $\nu$ we obtain:
\be 
\frac{m_{Pl}^2}{16\pi }\, R + \frac{1}{M^2}\left[ (R + 3 D^2) D_{\alpha} J^{ \alpha}_1 + J^{\alpha}_1 \,D_{\alpha}R  \right] - 
D_{\alpha} \phi \, D^{\alpha} \phi^* + 2 U(\phi ) = - \frac{1}{2} \, T_{\mu}^{\mu}\, .
\label{trace-eq}
\ee

The equation of motion for field $\phi$ is:
\be
D^2 \phi + \frac{\partial U}{\partial \phi^*} = - \frac{i q}{M^2} (2 D_{\mu} R\, D^{\mu} \phi + \phi D^2 R)\, .
\label{EoM-phi}
\ee

According to definition (\ref{current}), the current divergence is:
\be 
D_{\mu} J^{\mu}_1 =  \frac{2q^2}{M^2} \left[ D_{\mu} R\, (\phi^* D^{\mu}\phi + \phi D^{\mu}\phi^*) + |\phi|^2 D^2 R \right]
+ i q \left(\phi \frac{\partial U}{\partial \phi} - \phi^* \frac{\partial U}{\partial \phi^*} \right)\,.
\label{J-div}
\ee
If the potential of $U$ is invariant with respect to the phase rotation of $\phi$, i.e.
$U = U(|\phi|)$, the last term in this expression disappears. Still the current remains non-conserved, but this non-conservation
does not lead to any cosmological baryon asymmetry. Indeed, the current non-conservation is proportional to the product
$\phi^* \phi$, so it can produce or annihilate an equal number of baryons and antibaryons.

To create cosmological baryon asymmetry we need to introduce new types of interactions,
for example, the term in the potential of the form:
$ U_4 = \lambda_4 \phi^4 + \lambda_4^* \phi^{*4} $. This potential is surely non invariant w.r.t. the phase rotation of $\phi$
and can induce the B-non-conserving process of transition of two scalar baryons into two antibaryons,
$2 \phi \rar 2 \bar \phi$. An additional B-nonconserving interaction may contain some other fields, e.g. $\phi \bar q q$, 
where $q$ is a fermion, not necessarily a quark. Anyhow such new terms, or something more complicated,
can be included into the potential $U$. Of course, because of them the invariance of the 
theory with respect to the phase rotation of $\phi$ would be broken.

\subsection{ Current: version 2}

If $U = U(|\phi|)$ and so the theory is invariant with respect to the phase rotation, $\phi \rar \exp (i \beta) \phi$ with 
constant  $\beta$, then according to the Noether theorem there must exist the conserved current, which has the form 
\be
J_{2 \mu}  = i q \left( \phi^* \partial_\mu \phi -  \phi\, \partial_\mu \phi^* \right) -
\frac{2 q^2} {M^{2}}\, |\phi|^2 \,  D_\mu R\,.
\label{J-2}
\ee
We repeat that the last term in the above expression for the current is called the sea-gull term, the same as in scalar 
electrodynamics.

The equation of motion for the field $\phi$ is modified as:
\be
D^2 \phi + \frac{\partial U}{\partial \phi^*} = - \frac{i q}{M^2} (2 D_{\mu} R\, D^{\mu} \phi + \phi \,D^2 R)
+ \frac{2 q^2} {M^{4}} \,\phi \, D_\mu R\,D^\mu R \, ,
\label{EoM-phi-2}
\ee
and the current divergence now becomes:
\be 
D_{\mu} J^{\mu}_{2} =  i q \left(\phi \frac{\partial U}{\partial \phi} - \phi^* \frac{\partial U}{\partial \phi^*} \right)\,.
\label{J-div-2}
\ee
Let us repeat that  the r.h.s. of this equation vanishes, if $U = U(|\phi|)$.

The corresponding equations of motion for gravitational field acquire additional terms from the variation of the
sea-gull term in the current.  So they take the form:
\be \nonumber
&&\frac{m_{Pl}^2}{16\pi } \left( R_{\mu\nu} - \frac{1}{2} g_{\mu\nu} R \right)  \\ \nonumber
&& - \frac{1}{M^2}\left[ \left( R_{\mu \nu} - D_{\mu}D_{\nu} + g_{\mu\nu}D^2 \right) D_{\alpha} J^{\alpha}_{2} +
\frac{1}{2} g_{\mu\nu} J^{\alpha}_{2}\,D_{\alpha} R   - \frac{1}{2} (J_{2 \nu} D_{\mu} R + J_{2\mu}\, D_{\nu} R)\right]  \\ \nonumber
&& - \frac{2q^2}{M^4} \left( R_{\mu \nu} - D_{\mu}D_{\nu} + g_{\mu\nu}D^2 \right)  D_{\alpha} ( |\phi |^2\,D^{\alpha}R) \\ 
&& - \frac{1}{2} (D_{\mu} \phi \, D_{\nu} \phi^* + D_{\nu} \phi \, D_{\mu} \phi^*) + 
\frac{1}{2} g_{\mu\nu} \left[ D_{\alpha} \phi \, D^{\alpha} \phi^* -  U(\phi )\right] =  \frac{1}{2}\, T_{\mu\nu}\, .
\label{EoM-2}
\ee
We have verified that the covariant derivative $D_\mu$ acting on the
l.h.s. of Eqs.~(\ref{EoM}) or (\ref{EoM-2}) vanishes, as expected.

Taking the trace of equation (\ref{EoM-2}) with respect to $\mu$ and $\nu$ we obtain:
\be 
\frac{m_{Pl}^2}{16\pi }\, R + \frac{1}{M^2}\left[ (R + 3 D^2) D_{\alpha} J^{ \alpha}_2 + J^{\alpha}_2 \,D_{\alpha}R  \right] 
 + \frac{2q^2}{M^4} (R + 3 D^2) D_{\alpha}  ( |\phi |^2\,D^{\alpha}R) \\ \nonumber
- D_{\alpha} \phi \, D^{\alpha} \phi^* + 2 U(\phi ) = - \frac{1}{2} \, T_{\mu}^{\mu}\, .
\label{trace-eq-2}
\ee

\section{Solution in FRW background \label{sec-FRW}} 

Let us consider solutions of the above equations of motion in cosmology. The metric of the 
spatially flat cosmological FRW background can be taken as:
\be
ds^2=dt^2 - a^2(t) d{\bf r}^2\, .
\label{ds-2}
\ee

In the homogeneous case the  equation for the curvature scalar (\ref{trace-eq}) takes the form:
\be 
\frac{m_{Pl}^2}{16\pi }\, R + \frac{1}{M^2}\left[ (R + 3 \partial_t^2 + 9 H \partial_t) D_{\alpha} J^{\alpha}_1 + 
\dot R \, J_1^0 \right] 
= - \frac{ T^{(tot)}}{2} \,, 
\label{trace-eq-FRW}
\ee
where  $J_1^0$ is the baryonic number density of the $\phi$-field,
$H = \dot a/a$ is the Hubble parameter, and $T^{(tot)}$ is the trace of the energy-momentum tensor of matter including 
contribution from the $\phi$-field. In the homogeneous and isotropic cosmological plasma 
\be
T^{(tot)} = \rho - 3 P\, ,
\label{T-tot}
\ee
where $\rho$ and $P$ are respectively the energy density and the pressure of the plasma. 
For relativistic plasma $\rho = \pi^2 g_* T^4/30$ with $T$ and $g_*$ 
being respectively the plasma temperature and the number of particle species in the plasma. 
The Hubble parameter is expressed through $\rho$ as $H^2 = 8\pi \rho/(3m_{Pl}^2) \sim T^4/m_{Pl}^2$. 

The covariant divergence of the current is given by the expression (\ref{J-div}). In the
homogeneous case we are considering it takes the form:
\be 
D_{\alpha} J_1^{\alpha} = \frac{2q^2}{M^2} \left[ \dot R\, (\phi^* \dot \phi + \phi \dot \phi^*) + 
(\ddot R + 3H \dot R)\, \phi^*\phi\right]
+ i q \left(\phi \frac{\partial U}{\partial \phi} - \phi^* \frac{\partial U}{\partial \phi^*} \right)\,.
\label{J-div-hom}
\ee

To derive the equation of motion for the classical field $R$ in the cosmological plasma we have to take 
the expectation values of the products of the quantum operators $\phi$, $\phi^*$, and their derivatives. 
Performing  the thermal averaging,  as is described in the Appendix, we find 
\be
\langle \phi^* \phi \rangle =  \frac{T^2}{12}\, , \ \ \ 
\langle \phi^* \dot \phi + \dot \phi^* \phi \rangle = 0\, .
\label{av-phi}
\ee
Substituting these average values into eq. (\ref{trace-eq-FRW}) and neglecting the last term in 
eq.~(\ref{J-div-hom}) we obtain the fourth order differential equation:
\be
\frac{m_{Pl}^2}{16\pi }\, R + 
\frac{q^2}{6 M^4} \left(R + 3 \partial_t^2 + 9 H \partial_t \right)
\left[ 
\left(\ddot R + 3H  \dot R\right) T^2 \right] + 
\frac{1}{M^2} \dot R \, \langle J_1^0 \rangle 
= - \frac{T^{(tot)}}{2} \, .
\label{trace-eq-plasma}
\ee
Here $\langle J_1^0 \rangle $ is the thermal average value of the baryonic number density of $\phi$. It is assumed 
to be zero initially  and generated as a result of GBG. We neglect this term, since it is surely small initially and
probably subdominant later. Anyhow it does not noticeably change the exponential rise of $R$ at the onset of the
instability.

Eq.~(\ref{trace-eq-plasma}) can be further simplified if the variation of $R(t)$ is much faster than the universe expansion 
rate or in other words $\ddot R / \dot R \gg H$. Correspondingly the temperature may be considered adiabatically
constant. The validity of these assumption is justified a posteriori after we find  the solution for $R(t)$. 

Keeping only the linear in $R$ terms and neglecting higher powers of $R$, such as $R^2$ or $H R$, we obtain the
linear fourth order equation:
\be
\frac{d^4 R}{dt^4} + \mu^4 R =  - \frac{1}{2} \, T^{(tot)}\, . 
\label{d4-R}
\ee
where 
\be 
\mu^4 = \frac{m_{Pl}^2 M^4}{8 \pi q^2 T^2}\,.
\label{mu}
\ee

The homogeneous part of this equation has exponential solutions  $R \sim \exp (\lambda t)$ with
\be
\lambda = | \mu | \exp \left(  i\pi /4 + i \pi n /2 \right),
\label{lambda}
\ee
where $n = 0,1,2,3$. 

There are two solutions with positive real parts of $\lambda$.
This indicates that the curvature scalar is
exponentially unstable with respect to small perturbations, so $R$ should rise exponentially fast with time 
and quickly oscillate around this rising function.

Now we need to check if the characteristic rate of  the perturbation explosion is indeed much larger than the rate
of the universe expansion, that is:
\be 
(Re\, \lambda)^4   > H^4 = \left( \frac{ 8\pi \rho }{ 3 m^2_{Pl}}\right)^2 =  \frac{16 \pi^6 g_*^2}{2025}\,
\frac{ T^8}{m_{Pl}^4}, 
\label{lambda-to-H}
\ee
where $\rho = \pi^2 g_* T^4 /30$ is the energy density of the primeval plasma at temperature $T$
and $g_* \sim 10 - 100$ is the number of relativistic
degrees of freedom in the plasma. This condition is fulfilled if
\be
  \frac{2025}{2^9 \pi^7 q^2 g_*^2}\frac{m_{Pl}^6 M^4}{T^{10}} > 1\,,
\label{inst-OK}
\ee
or, roughly speaking, if $T \lesssim m_{Pl}^{3/5} M^{2/5} $.  Let us stress that at these temperatures the
instability is quickly developed and the standard cosmology would be destroyed.

If we want to preserve the successful  big bang nucleosynthesis (BBN)  results and  
impose the condition that the development of the instability 
was longer than the Hubble time at the BBN epoch at $T \sim 1 $ MeV, then $M$ should be extremely small,
$M < 10^{-32}$ MeV. The desire to keep the standard cosmology at smaller $T$ would demand even tinier $M$.
A tiny $M$ leads to a huge strength of coupling (\ref{L-GBG}). It surely would lead to pronounced effects in
stellar physics. 

\section{Discussion and conclusion \label{s-conclude}}

Thus we have shown that the curvature dependent addition (\ref{L-GBG}) to the Hilbert-Einstein Lagrangian leads to 
4th order differential equation of motion for the curvature scalar  (\ref{trace-eq-plasma}) or  (\ref{d4-R}). These 
equations are unstable with respect to small perturbations of the FRW background. The 
instability we discovered leads to
an exponential rise of the curvature. For a large range of cosmological temperatures the development of the instability 
is much faster than the universe expansion rate. The rise of $R$ could be terminated by the effects of non-linear
terms in the equations of motion. Evidently $R$ would stop rising if the non-linear terms become comparable by 
magnitude with the linear ones. It means that the rise terminates when $R$ by far exceeds the 
value found in the normal General Relativity.
So the simple version of GBG based on the coupling (\ref{L-GBG}) would be not compatible with observations and
some new stabilizing types of interaction are necessary.

These results are obtained for two possible forms of the scalar baryon currents, different by presence or 
absence of the sea-gull term, $|\phi|^2 \partial_\mu R $. Both forms of the current, $J_{1\mu}$ and $J_{2\mu}$,
can be used for the building of the GBG model. An important difference between these two currents is that in absence of 
B-non-conserving interactions of $\phi$ with other particles $J_{2\mu}$ is conserved,
while $J_{1\mu}$ is not. 

Lagrangian (\ref{L-GBG}) with current $J_2$ can be integrated by parts and rewritten
as $ - R D_\alpha J_2^\alpha/M^2 $. Naively it vanishes for conserved current, and does not lead to any observable 
effects. However, one should not use equations of motion in the Lagrangian. A well known example of that is the
Lagrangian for the Dirac fermions, which vanishes on the equations of motion.

In equation (\ref{trace-eq-plasma}) we have neglected the effects of matter and, in particular, of the
field $\phi$.
However, equations of motion for $\phi$ (\ref{EoM-phi}) and (\ref{EoM-phi-2})
demonstrate that evolution of $\phi$ depends upon $R$. Nevertheless, at the initial stage of instability when $R$ is not
yet huge, the impact of $R$ on $\phi$ may be neglected. At larger $R$ the effect may be essential and may lead to 
a noticeable impact on the baryogenesis.

One more comment is in order here. We assume that initially $R$ was close to its GR value, i.e.
$R = R_{GR} + R_1 (t) $, where $R_1$ is a small correction to $R_{GR}$ induced by the instability  
of eq.~(\ref{trace-eq-plasma}). So the background metric remains basically the usual FRW one
with slowly, or better to say, normally  changing cosmological scale factor. 
When and if $R_1$ becomes comparable by magnitude
to the canonical $R_{GR}$, the scale factor starts
to vary with the same speed as $R_1$  because of the relation (for spatially flat universe):
\be
R = - 6 \left(\frac{\ddot a}{a} + \frac{\dot a^2}{a^2}  \right).
\label{R-of-a}
\ee
When $R$ is stabilized, due to yet unknown mechanism, the universe quite probably would expand with acceleration. 
Interaction (\ref{L-GBG}) might mimic dark energy.

So, despite instability which drastically changes the outcome of gravitational baryogenesis, the scenario
with interaction (\ref{L-GBG}) is promising and may lead to interesting modifications of the standard cosmology.

\acknowledgments
Our work  was supported by the RSF Grant N 16-12-10037. 

\appendix

\section{Thermal averaging of the products of operators $\phi (x)$}
\label{s-averaging}

Field $\phi (x)$ can be expanded in terms of quantum creation-annihilation operators as follows:
\be
	\label{phi-x}
	\phi (x) = \int \frac{d^3q}{\sqrt{2 E_q\,(2\pi)^3 }}
	\left[ a({\bf q}) e^{- iE_q t + i {\bf q x}} + b^\dag ({\bf q}) e^{ iE_q t - i {\bf q x}} \right]\,,
\ee
where $E_q = \sqrt{q^2 + m^2_{\phi}}$ with $q=|\bf q|$. 
In eq. (\ref{phi-x}) $a$ and $a^{(\dag)}$, $b$ and  $b^{(\dag)}$ are the 
annihilation and creation operators for scalar  particles and antiparticles, respectively, 
obeying the commutation relations
\be
\left[a({\bf q}), a^\dag({\bf q'})\right]=2E_q\,(2 \pi)^3  \delta({\bf q}-{\bf q'})\,,
\label{commut}
\ee
the same for $b ({\bf q})$.

The products of creation-annihilation operators averaged over the medium have the 
standard form:
	\begin{eqnarray}
	\label{thermal_averages}
	\langle a^\dag({\bf q}) a({\bf q}') \rangle &=& 
f_B (E_q) \delta^{(3)} ({\bf q} - {\bf q}'),
	\hspace{1.5cm}
	\langle a({\bf q}) a^\dag({\bf q}') \rangle = [1 + f_B (E_q)] \delta^{(3)} 
({\bf q} - {\bf q}'),
	\cr\cr
	\langle b^\dag({\bf q}) b({\bf q}') \rangle &=& f_{\bar B} (E_q) \delta^{(3)} 
({\bf q} - {\bf q}'),
	\hspace{1.5cm}
	\langle b({\bf q}) b^\dag({\bf q}') \rangle = [1 + f_{\bar B} (E_q)] \delta^{(3)} 
({\bf q} - {\bf q}'),
	\end{eqnarray}
where $f_{B, \bar B}(E_q)$ is the energy dependent boson distribution function, 
which may be 
arbitrary since we assumed only that the medium is homogeneous and isotropic.
We have also assumed, as it is usually done, that non-diagonal matrix elements of
creation-annihilation operators vanish on the average due to decoherence.
For the vacuum state $f(E) = 0$ and we obtain the usual vacuum average values of 
$a a^\dag$ and $a^\dag a$, which  from now on will be neglected because we are 
interested only in the matter effects.

Averaging the product $ \phi^*(x) \phi(x) $ we obtain
\be \nonumber
\langle \phi^*(x) \phi(x) \rangle &=&  \\ \nonumber
&&
\int \frac{d^3q\,d^3q'}{2\, (2\pi)^3 \sqrt{E_q  E_{q'}}} 
\langle
[a^\dag({\bf q}) e^{ iE_q t - i {\bf q x}} + b ({\bf q}) e^{- iE_q t + i {\bf q x}} ]\,
		[a({\bf q'}) e^{- iE_{q'} t + i {\bf q' x}} + b^\dag ({\bf q'}) e^{ iE_{q'} t - i {\bf q' x}} ]
\rangle	\\ 
&&= \int \frac{d^3q}{(2\pi)^3 E_q} f(E_q, T)\,. 
\label{phi-averaging}
\ee  

In thermal equilibrium the distributions of bosons  and their antiparticles
with zero chemical potentials 
have the usual Bose-Einstein  form:
	\begin{eqnarray}
	\label{No_Bose_defs}
	f_B(E,T) =   f_{\bar B} (E,T) = \frac{1}{\exp (E/T) - 1},
		\end{eqnarray}
where in high temperature limit we can neglect the particle mass i.e. we can assume 
$E_q = q$.

The last integral in Eq.~(\ref{phi-averaging}) is simply taken giving:
\be 
\langle \phi^*(x) \phi(x) \rangle = \frac{1}{2 \pi^2} \int \frac{dq\, q}{e^{q/T} - 1} = 
\frac{T^2}{2 \pi^2} \int_{0}^{\infty} \frac{ dz\,z}{e^z - 1} = \frac{T^2}{12}.
\label{aver-int}
\ee 

The average value of the operator $\langle \phi^* \dot \phi + \dot \phi^* \phi \rangle$ is zero.

\end{document}